\documentclass{article}

\usepackage[square,numbers]{natbib}

\usepackage[final]{neurips_2023_ml4ps}

\usepackage{amsmath}
\usepackage{bm}
\usepackage{amssymb}

\usepackage[utf8]{inputenc} 
\usepackage[T1]{fontenc}    
\usepackage{hyperref}       
\usepackage{url}            
\usepackage{booktabs}       
\usepackage{amsfonts}       
\usepackage{nicefrac}       
\usepackage{microtype}      
\usepackage{xcolor}         
\usepackage{subfiles}
\usepackage{amsmath} 
\usepackage{graphicx}
\usepackage{bm}
\usepackage{caption}
\usepackage{subcaption}
\usepackage{float}

\title{Accelerating Multiphase Flow Simulations with Denoising Diffusion Model Driven Initializations}

\author{%
Jaehong Chung$^{1, 2}$ \quad Agnese Marcato$^2$ \quad Eric J. Guiltinan$^2$ \\
\textbf{Tapan Mukerji}$^1$ \quad \textbf{Hari Viswanathan}$^2$ \textbf{Yen Ting Lin}$^2$ \quad \textbf{Javier E. Santos}$^2$\\\\
$^1$Stanford University \quad $^2$Los Alamos National Laboratory\\
\\
\\
}

\begin{document}
\maketitle

\begin{abstract}

This study introduces a hybrid fluid simulation approach that integrates generative diffusion models with physics-based simulations, aiming at reducing the computational costs of flow simulations while still honoring all the physical properties of interest. These simulations enhance our understanding of applications such as assessing hydrogen and CO$_2$ storage efficiency in underground reservoirs. Nevertheless, they are computationally expensive and the presence of nonunique solutions can require multiple simulations within a single geometry. To overcome the computational cost hurdle, we propose a hybrid method that couples generative diffusion models and physics-based modeling. We introduce a system to condition the diffusion model with a geometry of interest, allowing to produce variable fluid saturations in the same geometry. While training the model, we simultaneously generate initial conditions and perform physics-based simulations using these conditions. This integrated approach enables us to receive real-time feedback on a single compute node equipped with both CPUs and GPUs. By efficiently managing these processes within one compute node, we can continuously evaluate performance and stop training when the desired criteria are met. To test our model, we generate realizations in a real Berea sandstone fracture which shows that our technique is up to 4.4 times faster than commonly used flow simulation initializations.
\end{abstract}

\section{Introduction}

CO$_2$ and H$_2$ subsurface storage are seen as important methods to address climate change \cite{dai2016co2, krevor2023subsurface}. The injection of CO$_2$ and H$_2$ is a multiphase process whereby these fluids displace existing groundwater. Predicting the behavior of these two-phase fluid systems is challenging, as their dynamics are governed by diverse factors including flow path geometries \cite{blunt2001flow}, fluid saturations \cite{parker1989multiphase}, and affinity between the fluids and the host rocks \cite{fatt1959effect}. In addition, the presence of fractures, which act as preferential pathways for flow, further complicates the picture \cite{viswanathan2022fluid}. These fractures are potential risks for fluid leakage \cite{tongwa2013evaluation}, underlining the need for a comprehensive understanding to ensure the integrity and efficiency of underground storage systems \cite{fitts2013caprock, guiltinan2021two, ting2022using}.

Pore-scale simulations, ranging from 10 nm to 10 cm, provide accurate models of how fluids travel through porous media in underground reservoirs \cite{middleton2012cross}. In particular, these micro-scale simulations can capture the preferred traveling paths for fluids by identifying local minimum energy states \cite{blunt2013pore}. Such configurations arise from the energy balance between fluid displacements and interfacial tensions between fluids and solids \cite{blunt2017multiphase}. While the pore-scale simulations can characterize accurately and precisely fluid configurations in the pore space, scaling these simulations to the reservoir scale, ranging from $10 \, cm$ to $100 \, m$, is impractical due to the expensive computational costs associated with large modeling domains with intricate flow paths (complicated geometries). In addition, the need to run multiple simulations for different saturation levels in the same geometries (for example, to compute relative permeability curves) further escalates the computational challenges.

Many studies employ machine learning techniques for predicting fluid flow at the macroscale \cite{yan2022physics, yan2022gradient}. However, applications at the pore scale remain limited. Recent advancements in machine learning have demonstrated significant potential in reducing the computational burden of multiphase flow simulations. Machine learning can help these simulations by offering efficient algorithms that can learn complex patterns and relationships within the data. This capability allows for faster and more accurate predictions without the need for solving extensive and computationally expensive equations traditionally required in multiphase flow modeling. Additionally, machine learning models can be trained on large datasets to generalize well to new, unseen data, thereby enhancing the overall efficiency and scalability of the simulations.

Building on these advancements, several recent studies have explored the application of machine learning techniques to multiphase flow problems with encouraging results.
\cite{zhao2023fast} employed a U-Net architecture to predict fluid displacement in micromodels showing high accuracy in the predicted displacement configurations. 
\cite{ting2022using} trained a residual U-Net to predict two-phase configurations in fractures, mapping the 3D geometry into a 2D feature space with a lossless algorithm. Their trained model was able to predict the invasion dynamic of an unsteady state flow of a non-wetting front. 
\cite{guiltinan2020residual} trained a network to predict the fluid distribution within fractures at steady state using data from lattice Boltzmann simulations. They demonstrated that a trained network can accurately predict fluid residual saturation and distribution based solely on the dry fracture characteristics. 
\cite{wang2022poreA} trained a conditional Generative Adversarial Network to predict invasion percolation configurations in 2D spherepacks. While the model demonstrated a reduction in computational time compared to traditional algorithms, its use cases are still limited as these simulations do not capture the full complexity of two-phase fluid interactions. These studies provide an initial demonstration of successful applications of machine learning for multiphase flow problems.

Hybrid approaches, which combine the computational efficiency of deep learning with the physical robustness of numerical simulations, have been proposed as a solution to simulate complex physical problems efficiently. \cite{wang2021ml} introduced an integrated framework that employed a Convolutional Neural Network (CNN) to provide a data-driven initialization, subsequently utilized in single-phase Lattice Boltzmann Method (LBM) simulations. This approach achieved a tenfold acceleration of the simulation process while maintaining physical accuracy. In addition, \cite{chang2022improving} demonstrated that machine learning-based initializations can greatly accelerate simulations of electrical conductivity in porous media. These studies show the potential of a hybrid approach to combine the strengths of deep learning and numerical solvers. Our study aims to extend this hybrid approach to multiphase flow simulations, which are orders of magnitude more expensive compared to single phase and electrical conductivity simulations. Hybrid approaches tackle the challenge of achieving computational efficiency while ensuring compliance with fundamental governing equations, such as continuity and momentum balance.

In deep learning, generative models are designed to approximate high-dimensional data distributions \cite{goodfellow}. Recently, diffusion models have gained prominence for their unique ability to progressively convert noise into highly detailed and structured outputs that approximate the distribution of the training data. These models employ a reverse diffusion process to gradually build coherent structures from noise, and a forward diffusion process to incrementally corrupt the data. This bidirectional approach makes diffusion models particularly effective for complex computer vision tasks, such as generating high-resolution, realistic images \cite{ho2020denoising, sohl2015deep}, establishing new state-of-the-art for generative modeling.


In this study, we demonstrate that diffusion models can learn to accurately predict steady-state multiphase fluid configurations in fracture geometries. We introduce a geometric conditioning approach to obtain fluid configurations for specific fracture geometries. The predicted fluid configurations are highly accurate; we used them as initial conditions for physics-based simulations, significantly reducing computational time. We quantify accuracy using both statistical metrics, such as mean squared error, and by measuring the reduction in simulation time, indicating how closely the predicted configuration matches the real one. Our results demonstrate that diffusion models can provide accurate configurations for multiphase flow, significantly enhancing the efficiency of physics-based simulations. This opens the door to hybrid approaches where machine learning and physics-based methods can coexist, leveraging the strengths of both to solve complex problems more effectively.

The remainder of this paper is organized as follows: Section \ref{sec:Data} provides detailed information about the fracture dataset and multiphase simulations used in this study. Section \ref{sec:Method} describes the diffusion model used and details the geometric conditioning process for specific fracture geometries. Finally, Section \ref{sec:Results} presents and discusses the results. 
\section{Dataset}
\label{sec:Data}
\subsection{Fracture geometries}
\begin{figure}[h!]
  \centering
  \includegraphics[width=1\linewidth]{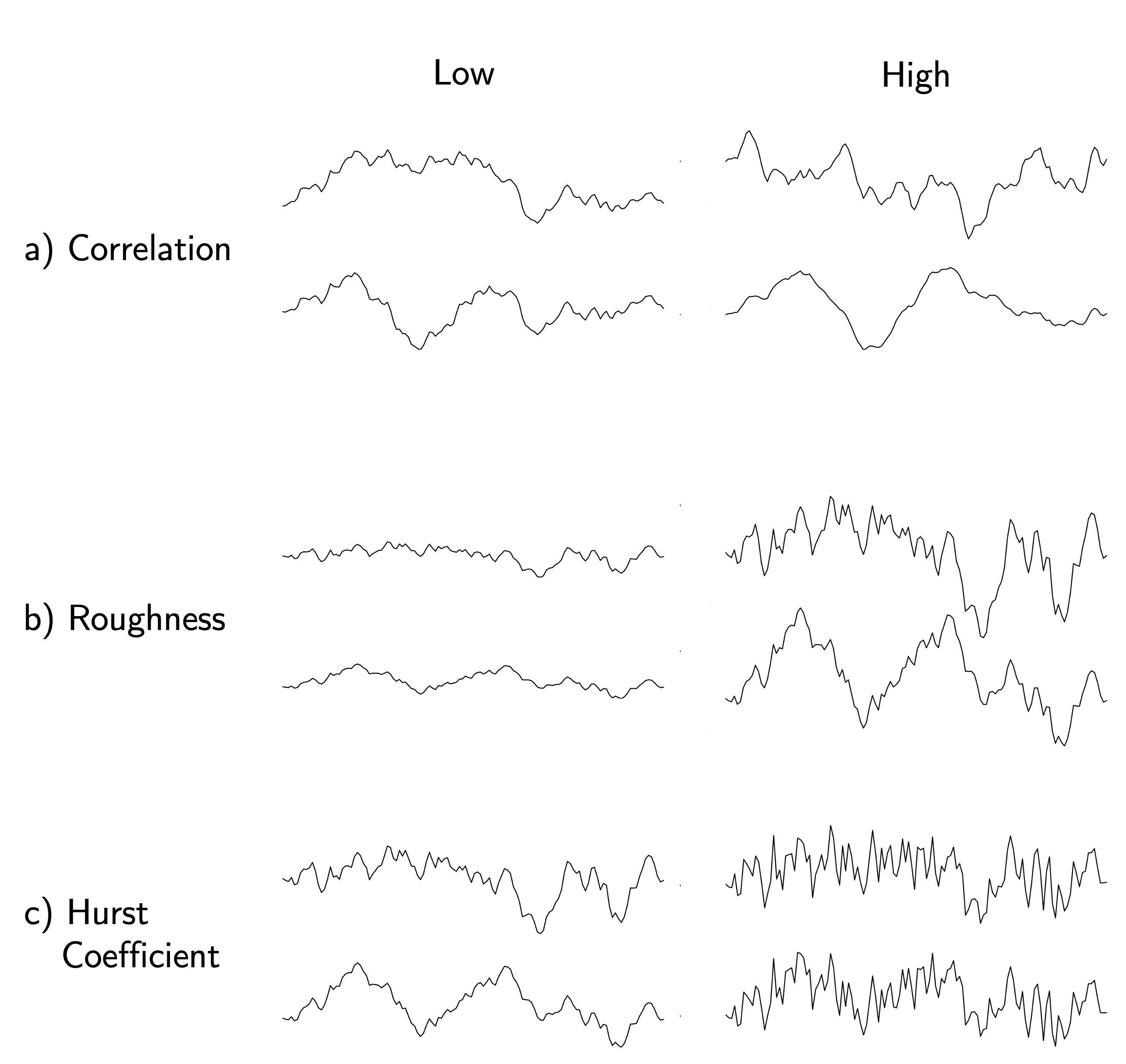}
  \caption{ Impact of geometric parameters on fracture characteristics: (a) Increasing spatial correlation in the flow direction, (b) Pronounced peaks and valleys with increased roughness, (c) Higher Hurst coefficient reduces local correlation, making individual asperities more pronounced while maintaining constant overall mean aperture. }
  \label{fig:parameters-effect}    
\end{figure}

To train our model, we generated a dataset of 1,000 fractures using \texttt{pySimFrac}, an open-source Python-based fracture geometry generator \cite{guiltinan2023pysimfrac}. We used the spectral method to create fractures by varying the following geometric parameters: Hurst exponent,  spatial correlation, and surface roughness. The impact of these parameters is illustrated in Figure \ref{fig:parameters-effect}.


Although these fractures are synthetic, the \texttt{pySimFrac} implementation and values used align with those measured in profilometry studies of real fractures \cite{brown1995simple}. For readers interested in the underlying methods and their implementations, further details can be found in the literature \cite{ glover1998synthetic, ogilvie2006fluid, guiltinan2023pysimfrac}.

\begin{figure}[h!]
  \centering
  \includegraphics[width=1\linewidth]{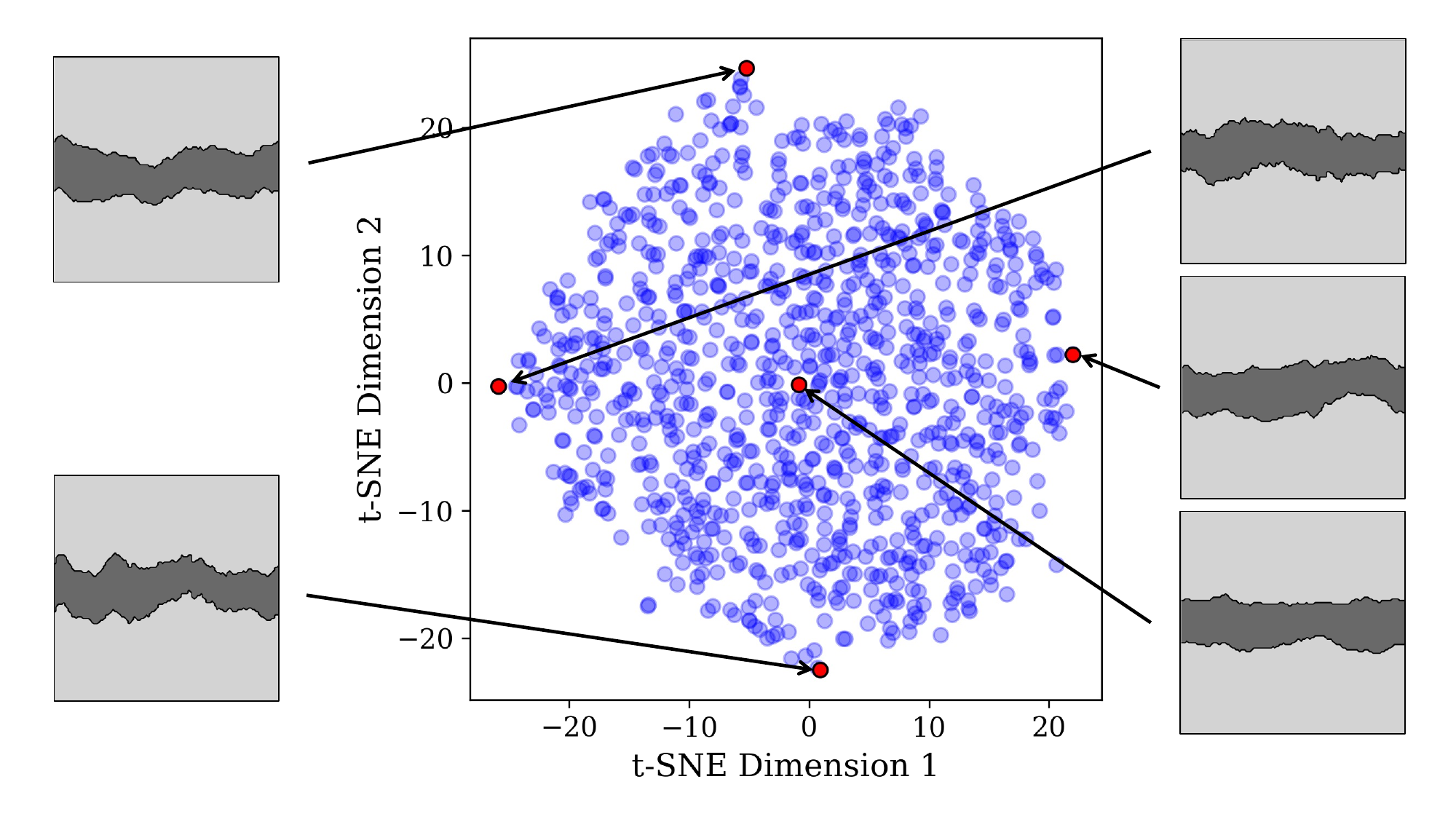}

  \caption{This t-SNE visualization reduces our synthetic rough fracture dataset into a two-dimensional space, condensing the original fracture geometry (16,384 pixels) into two dimensions (Dimension 1 and Dimension 2). The display reveals a continuous spectrum of data rather than isolated groups or clusters, indicating a diverse yet evenly distributed dataset. Several sets of hyperparameters were tested, consistently resulting in a similar distribution of points without significant clustering. We highlight a few examples to showcase the range of fractures in the dataset.}
  \label{fig:t-SNE}    
\end{figure}

The generated geometries are projected onto a 128-by-128 grid, enabling us to run numerous lattice Boltzmann simulations within a computationally feasible time frame. This approach enables us to create a sufficiently diverse dataset for training machine learning models. To visualize the diversity within our dataset, we employed t-SNE (t-Distributed Stochastic Neighbor Embedding), a technique for reducing dimensionality and visualizing high-dimensional data structures in a lower-dimensional space \cite{van2008visualizing}. Figure \ref{fig:t-SNE} shows that while differences exist within our dataset, the extremes of the t-SNE plot are relatively limited.  The goal of this work is to demonstrate the ability of the machine learning model to generalize to a variety of fractures despite the simplicity of its training set.


\subsection{Multiphase flow simulations}
We conducted two-phase flow simulations on the fractures geometries using the MP-LBM library \cite{santos2022mplbm} which uses the lattice Boltzmann method (LBM) to simulate two competing immiscible (wetting and non-wetting) fluids in complex geometries. In our study, the density distribution function for each phase is represented by the Boltzmann equation with the Bhatnagar-Gross-Krook (BGK) collision term \cite{bhatnagar1954model} as follows:

\begin{equation}
    f_a^{\alpha}\left(\bm{x}+\bm{e}_a\Delta t,t+\Delta t\right) = f_a^{\alpha}\left(\bm{x}, t\right) - \frac{1}{\tau}\left[f_a^{\alpha}\left(\bm{x}, t\right)-f_a^{\alpha, eq}\left(\bm{x}, t\right)\right],
\end{equation}
where $f_a^{\alpha}(\bm{x}, t)$ is the density function with the $a$th lattice direction, $\alpha$ represents components for two phase systems $\alpha = 1$ or $-1$, $\bm{e}_a$ is the set of discrete velocity vectors of a node, and $\tau$ is characteristic relaxation time, related to the fluid viscosity. The equilibrium distribution function, $f_a^{\alpha, eq}(\bm{x}, t)$, describes the distribution function in a state of local equilibrium \cite{chen1992lattice}. In addition, we consider the acting force on each fluid phase due to the fluid-fluid interaction (interfacial tension) and fluid-solid interaction (wettability) based on the Shan-Chen model \cite{shan1993lattice}. Readers interested in the underlying model for specifying a specific contact angle can refer to the literature \cite{huang2007proposed}.

The simulations were conducted using a 1:1 density ratio and a wetting angle of 24 degrees, representing a system with brine and supercritical CO2. It's worth noting that the proposed workflow is capable of learning systems with different fluid properties or wetting angles, it is not limited to our specific system.For each fracture, we ran thirteen evenly distributed simulations with fluid ratios varying from 20\% to 60\% to ensure percolation of both fluids and to exclude isolated blobs that would not converge in the simulation. This resulted in a total of 13,000 cases (1,000 synthetic fractures $\times$ 13 saturation cases) for the fluid configuration dataset in this study.

During the simulation, both fluids are driven through the fracture by an external force. Convergence is monitored by tracking the relative kinetic energy ($E_k$), which is represented as follows \cite{latt2021palabos}:
\begin{equation}
     E_k = \frac{1}{2} \left( \frac{\partial \vec{v}}{\partial t} \right)^2,
\end{equation}
where ${\partial \vec{v}}/{\partial t}$ denotes the time rate of change of the velocity vector. The simulation is considered converged once the relative energy change across the simulation domain falls below a threshold of $5 \times 10^{-5}$ between consecutive 1,000 time steps. The simulations take on average around 100 thousand iterations in time (time-steps) to achieve convergence. Once the simulation converges, the fracture's fluid configurations can be considered stable, implying a minimum free energy state. In other words, the converged simulation depicts the most conductive pathways for each sample given the solid geometry and the initial ratios of fluid. It is noteworthy that these stable fluid configurations exhibit intricate variability, even within the same fracture geometry.

In this work, the LBM simulations act as the ground-truth to train our model. Nevertheless, we would like to emphasize that the workflow described above is not limited to LBM simulations, it can also be applied using other methods such as level-set simulations \cite{jettestuen2013level}, molecular dynamics \cite{santos2020modeling}, and computational fluid dynamics\cite{bouras2022use}, among others. We chose LBM simulations because it is practical for us, given that we already have a robust open-source solver in place.


\section{Diffusion Models conditioned on fracture geometries}
\label{sec:Method}

We employ Denoising Diffusion Probabilistic Models \cite{ho2020denoising} to learn data distribution of stable fluid configurations in fractured media. These models operate by constructing two Markov Chains. A forward Markov chain is defined to perturb samples drawn from the data distribution, \(\mathbf{x_0} \sim q(\mathbf{x_0})\), towards a limiting isotropic Gaussian distribution. The conditional mean of the reverse-time process was derived \cite{sohl2015deep,ho2020denoising} as the training target of a U-Net \cite{ronneberger2015u}, which is used to propagate samples drawn from the limiting distribution back to the data distribution after training, generating novel samples. 
In short, the reverse diffusion process progressively transforms noise into coherent structures, while the forward diffusion process destroys the structure in the data distribution incrementally. The forward and the reverse processes are defined as follows~\cite{ho2020denoising}:
\begin{align}
    q(\mathbf{x_t}|\mathbf{x_{t-1}}) &= \mathcal{N}(\mathbf{x_t}; \sqrt{1-\beta_t}\mathbf{x_{t-1}}, \beta_t \mathbf{I}),\\
    p_\theta(\mathbf{x_{t-1}}|\mathbf{x_t}) &= \mathcal{N}(\mathbf{x_{t-1}}; \mu_{\theta} (\mathbf{x_t}, t), \Sigma_\theta(\mathbf{x_t}, t)),
\end{align}
where $\bm{x}_t$ is the sample at time $t$, $\beta_t$ is a variance schedule controlling the Gaussian noise addition, $\mu$ is a mean, and $\Sigma$ is a covariance matrix. 

Variational inference (VI) is used to train the diffusion model via minimizing the evidence lower bound (ELBO), a lower bound of the Bayes evidence of the model \cite{sohl2015deep,ho2020denoising}. \cite{ho2020denoising} showed that VI can be effectively performed by training the model to predict the noise, $\bm{\epsilon}$, that is added to $\bm{x}_t$, given a noisy image $\bm{x}_t$, via the loss function: 
\begin{equation}
    \mathcal{L}(\theta) = \mathbb{E}_{t \sim [1,T], \bm{x}_0, \bm{\epsilon}} [\| \bm{\epsilon} - \bm{\epsilon}_\theta (\bm{x}_t, t) \|^{2}],
\end{equation}

\subsection{Fracture geometry conditioning}

Since the geometry of flow paths has a first order influence on the configurations of fluids \cite{blunt2017multiphase}, a useful model should be able to provide fluid configurations for user-specified geometries. This task is similar to inpainting in computer vision \cite{yeh2017semantic, yu2018generative}, which requires the model to synthesize images in unknown areas that are both realistic and consistent with the surrounding background. One unique challenge in our inpainting task lies in the fracture geometries which are characterized by sharp and irregular shapes. To ensure that our model generates fluid configurations that adhere to given geometries (i.e., not solely based on random Gaussian noise), we introduce an auxiliary binary image $\bm{G} \in \{0,1\}^{H \times W}$ encoding the fracture geometry. Here, $H$ is the height and $W$ is the width of the image:
\begin{equation}
    \boldsymbol{G}(i,j) = 
    \left\{
    \begin{array}{ll}
    1 & \mbox{if pixel at } (i,j) \mbox{ is part of the solid}, \\
    0 & \mbox{otherwise}.
    \end{array}
    \right.
\end{equation}

\begin{figure}[h!]
    \centering
    \includegraphics[width=\textwidth]{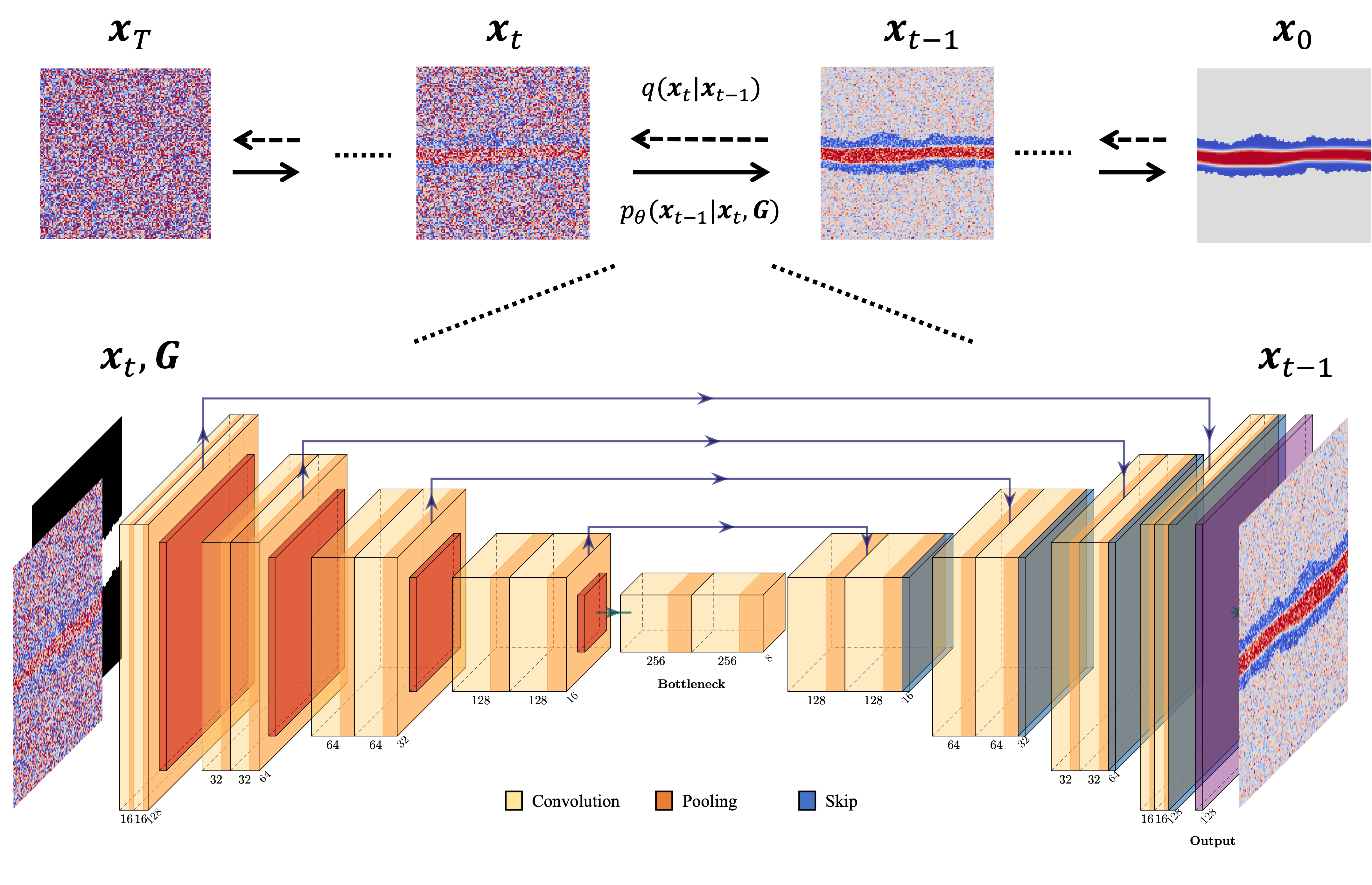}
    \caption{Diagram depicting the diffusion process and the integration of a geometry channel to guide fluid configurations within predefined geometric constraints.}
    \label{fig:diffusion_process}
\end{figure}

Figure  \ref{fig:diffusion_process}  shows a schematic diagram of the diffusion process in our model, which generates denoised images ($\bm{x}_{t-1}$) given $\bm{x}_t$ \emph{and} $\bm{G}$. By including this additional channel in every denoising step during the reverse process, the model is provided with essential information to synthesize images based on a given geometry. Once trained, the model can generate fluid realizations for unseen geometries by concatenating the desired fracture geometry, represented as a binary image  ($\bm{G}$), with a random isotropic Gaussian field ($\bm{x}_T$), and running the reverse process depicted in Figure \ref{fig:diffusion_process}. 

During the model's development, we experimented with using only one channel, where the solid regions were depicted by zeros, and the noise and denoising processes occurred within the fracture space. This approach resulted in very poor outcomes. We hypothesize that in highly rough fractures with very small pore spaces, the limited amount of Gaussian noise added is insufficient to enable the necessary expressiveness for the diffusion process to accurately create fluid configurations.

\section{Results and Discussion}
\label{sec:Results}
\subsection{Training and model performance}
We train the diffusion model with 100 denoising steps over 4,000 training iterations, a process that takes approximately one hour on a single A100 GPU. During training, we employ two evaluation metrics to assess the model's performance. The first metric is the Mean Squared Error (MSE), commonly used to train machine learning models, which quantifies the discrepancy between predictions and ground truth. While a low MSE value indicates that the model is effectively approximating the training data, determining the optimal model for the physical domain remains challenging.

To address this challenge, we introduce a second metric that focuses on the number of iterations required for the LBM simulation to converge, starting from the initial configuration provided by the diffusion model. The rationale behind this metric is based on the functionality of numerical simulators, which approximate solutions to partial differential equations through iterative methods. Therefore, if the fluid configuration generated by our diffusion model closely approximates the numerical solution, the number of iterations needed for convergence should be minimal, as the starting point is already near the converged configuration.

\begin{figure}[ht]
    \centering
    \begin{minipage}{0.48\textwidth}
        \includegraphics[width=\textwidth]{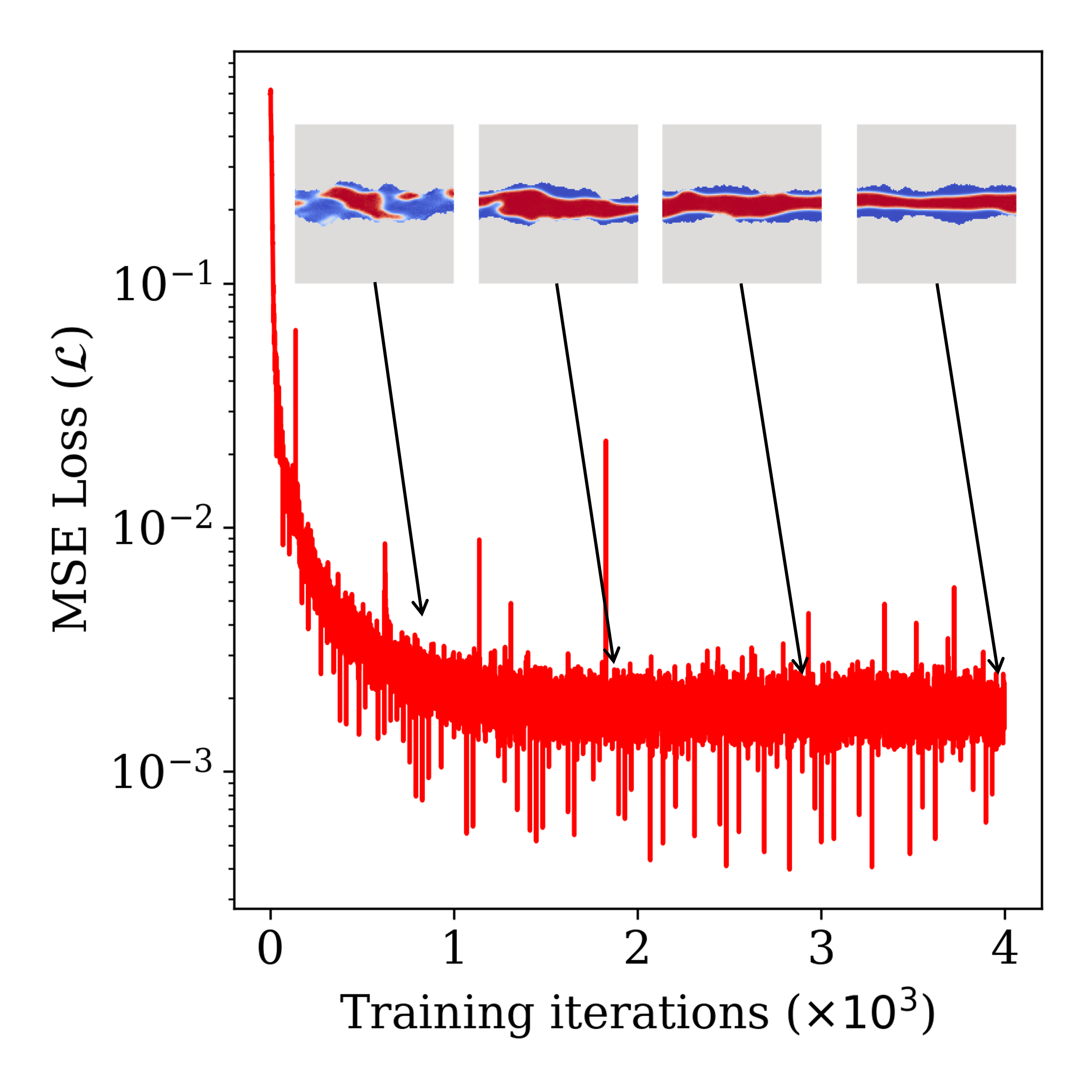}
        \centering
    \end{minipage}
    \hfill
    \begin{minipage}{0.48\textwidth}
        \includegraphics[width=\textwidth]{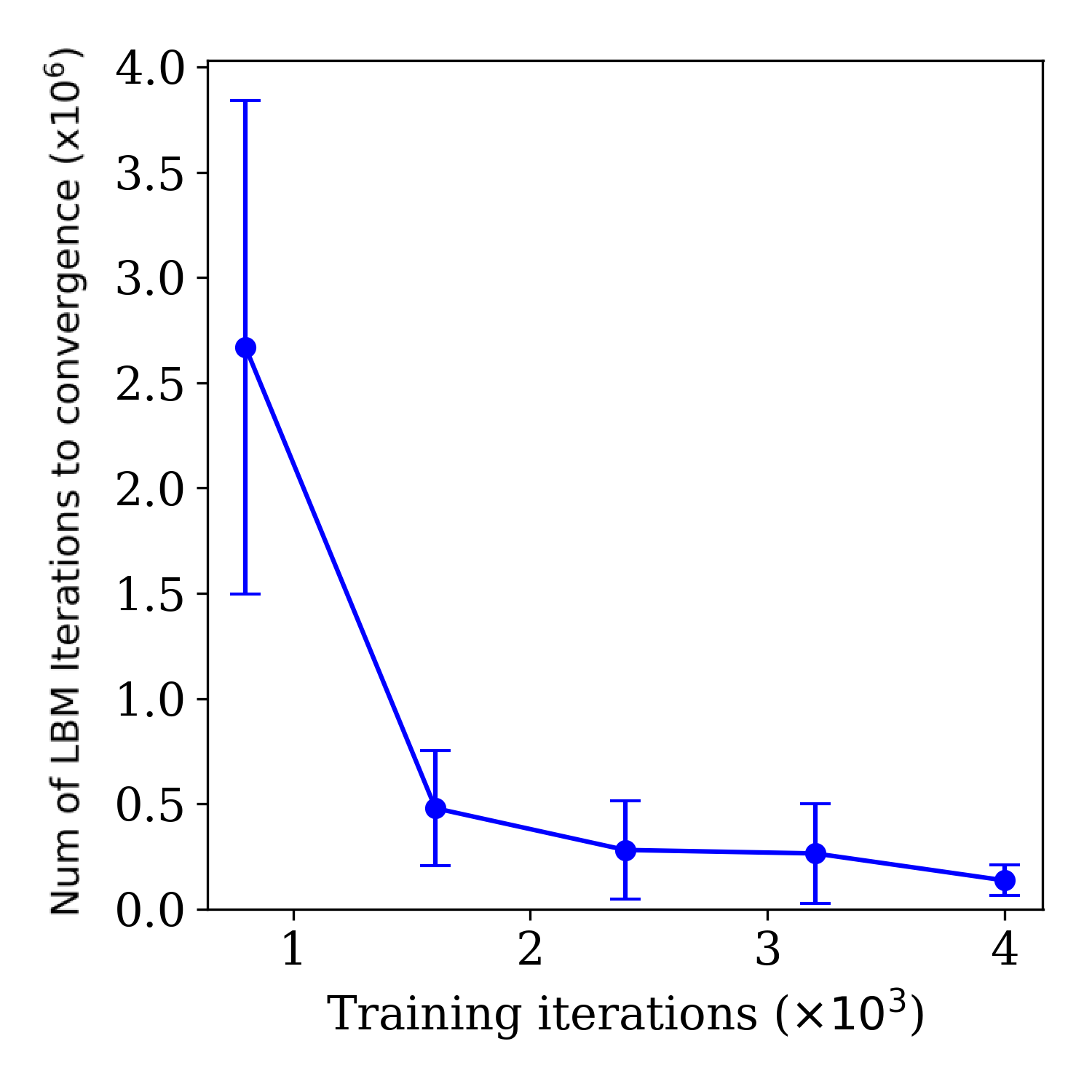}
        \centering
    \end{minipage}
    \caption{Evaluation metrics during diffusion model training: Left - Evolution of MSE loss across iterations, with highlighted samples at 1K, 2K, 3K, and 4K training iterations. Right - Variation in iterations required for convergence, showing the mean and standard deviation.}
    \label{fig:subplot_comparison}
\end{figure}

Throughout the training, we generate fluid configurations for unseen fracture geometries in intervals of 800 training iterations (during which the model has seen 800 fractures with varying noise levels). These configurations serve as initial conditions for multiphase flow simulations, and we record the number of iterations required by our LBM solver to achieve convergence. Figure \ref{fig:subplot_comparison}  illustrates the evolution of the MSE loss and the convergence iterations over the training period. Notably, the MSE loss starts high and plateaus around 2,000 training iterations, indicating effective denoising for the training dataset. Additionally, we observe a significant reduction in the number of iterations required for convergence.  Specifically, the average number of iterations decreased from 2.6 million to 140 thousand, representing a reduction of approximately 95\%. The standard deviation also decreased by 95\%, further highlighting the model's increasing consistency. These results suggest that our diffusion model not only captures the visual similarity of the training dataset but also learns to generate physical configurations conditioned on the fracture geometry. It is important to note that computing this metric (and running the multiphase lattice Boltzmann simulations) does not incur additional computational overhead, as the simulations are run on the fly using the idle CPUs of our compute node. This approach maximizes the utilization of our computing infrastructure and allows for continuous monitoring of the model's performance without additional resource demands.

\subsection{Assessing the Trained Model's Performance Against Different Initialization Methods}

\begin{figure}[h]
    \centering
    \begin{minipage}{0.20\textwidth}
        \includegraphics[width=\textwidth]{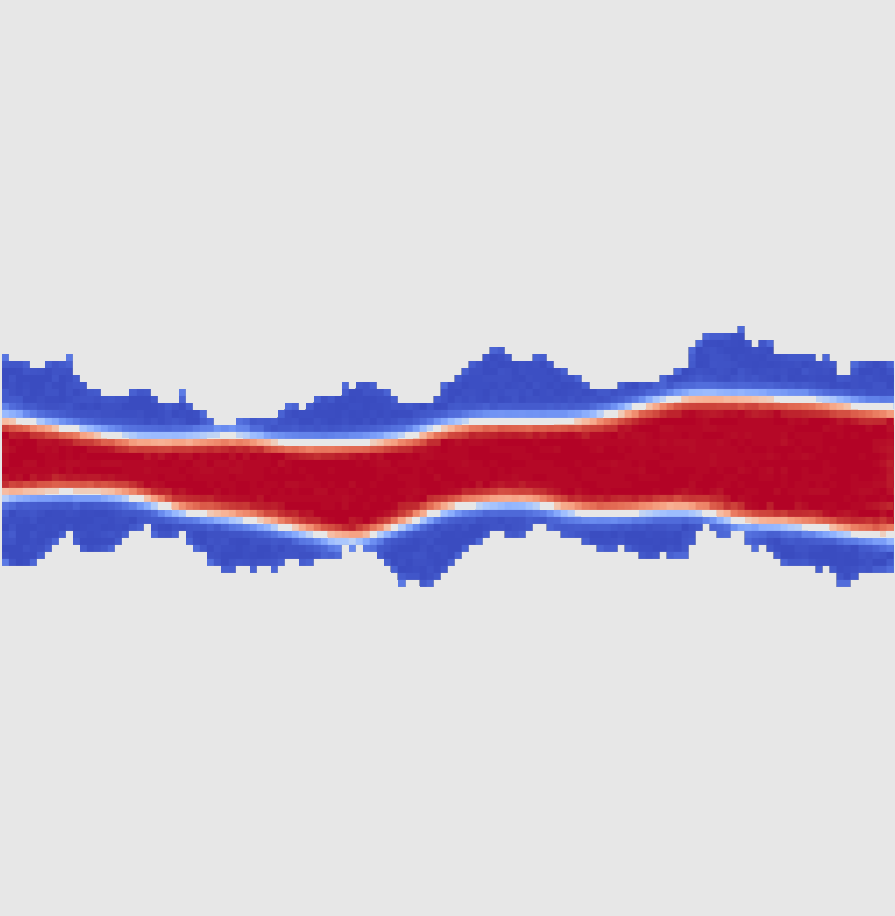}
        \centering
        \text{(a) Diffusion}
    \end{minipage}
    \hfill
    \begin{minipage}{0.20\textwidth}
        \includegraphics[width=\textwidth]{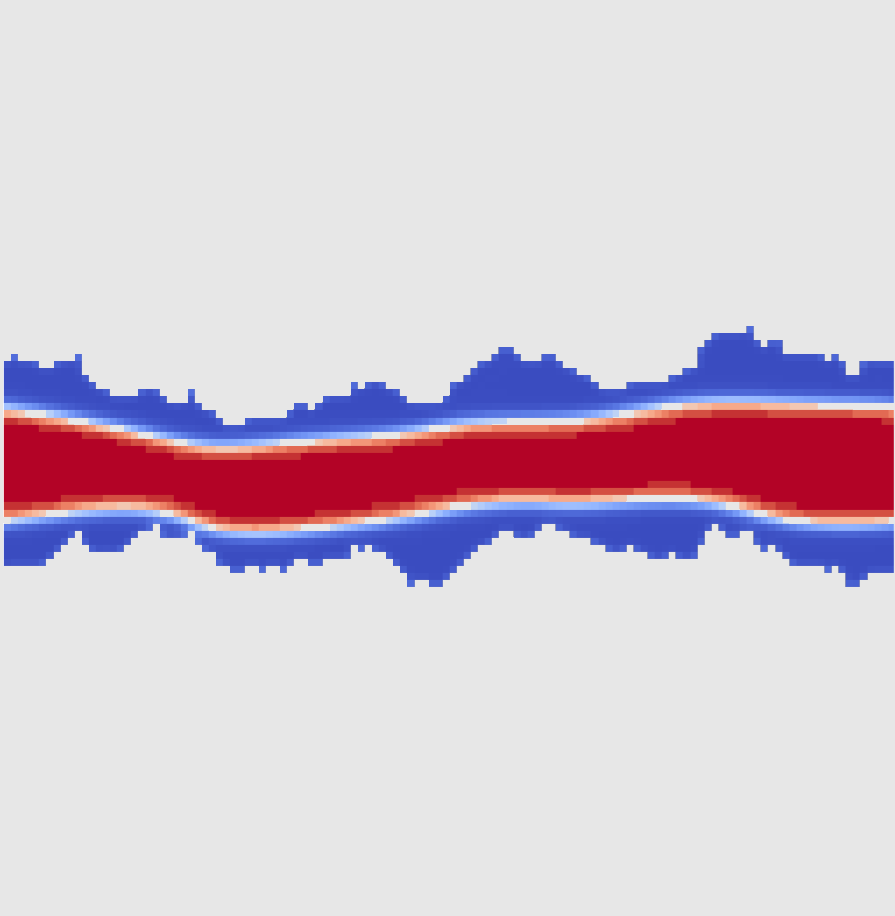}
        \centering
        \text{(b) Sim. solution}
    \end{minipage}
    \hfill
    \begin{minipage}{0.20\textwidth}
        \includegraphics[width=\textwidth]{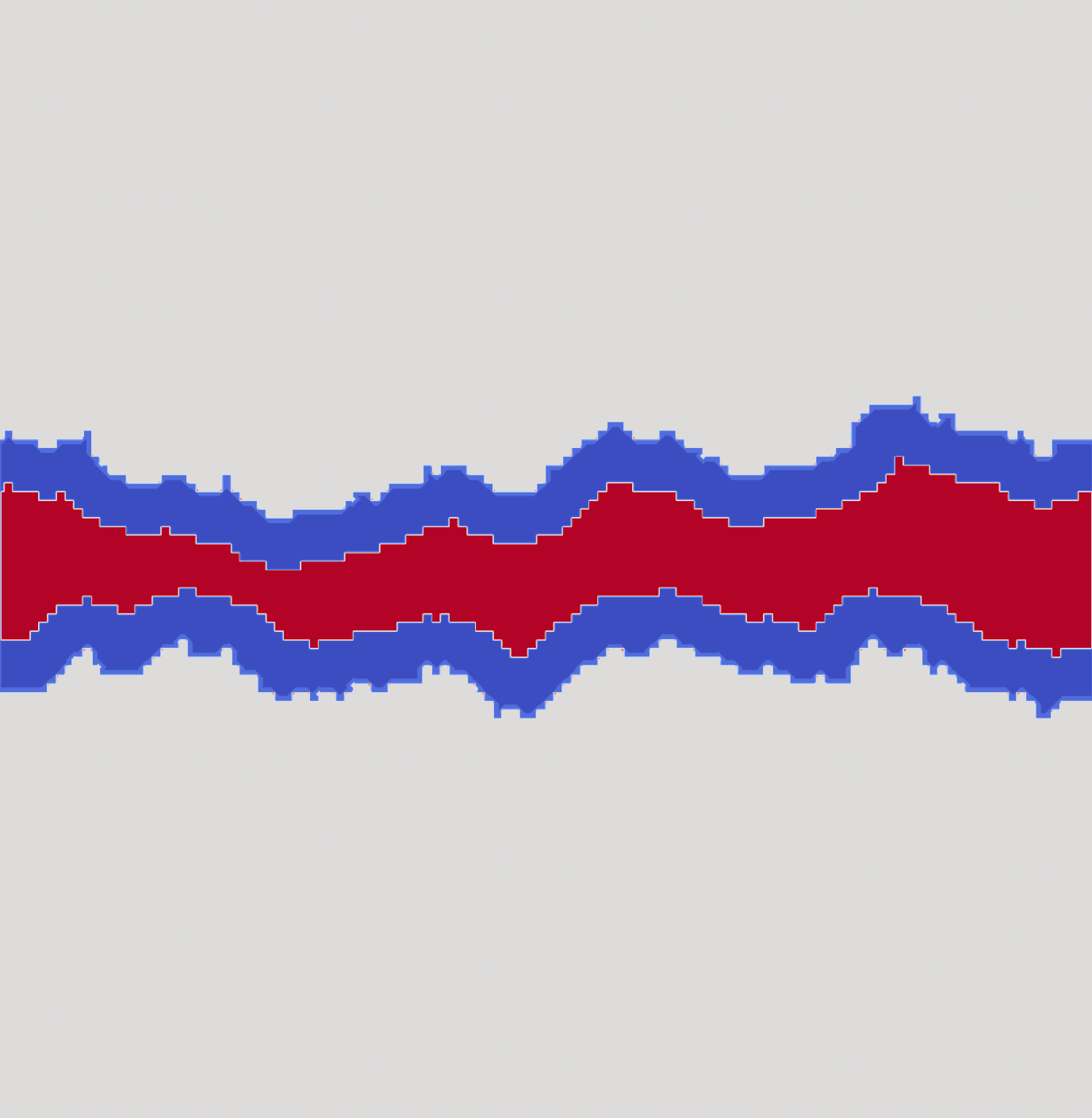}
        \centering
        \text{(c) Euclidean}
    \end{minipage}
    \hfill
    \begin{minipage}{0.20\textwidth}
        \includegraphics[width=\textwidth]{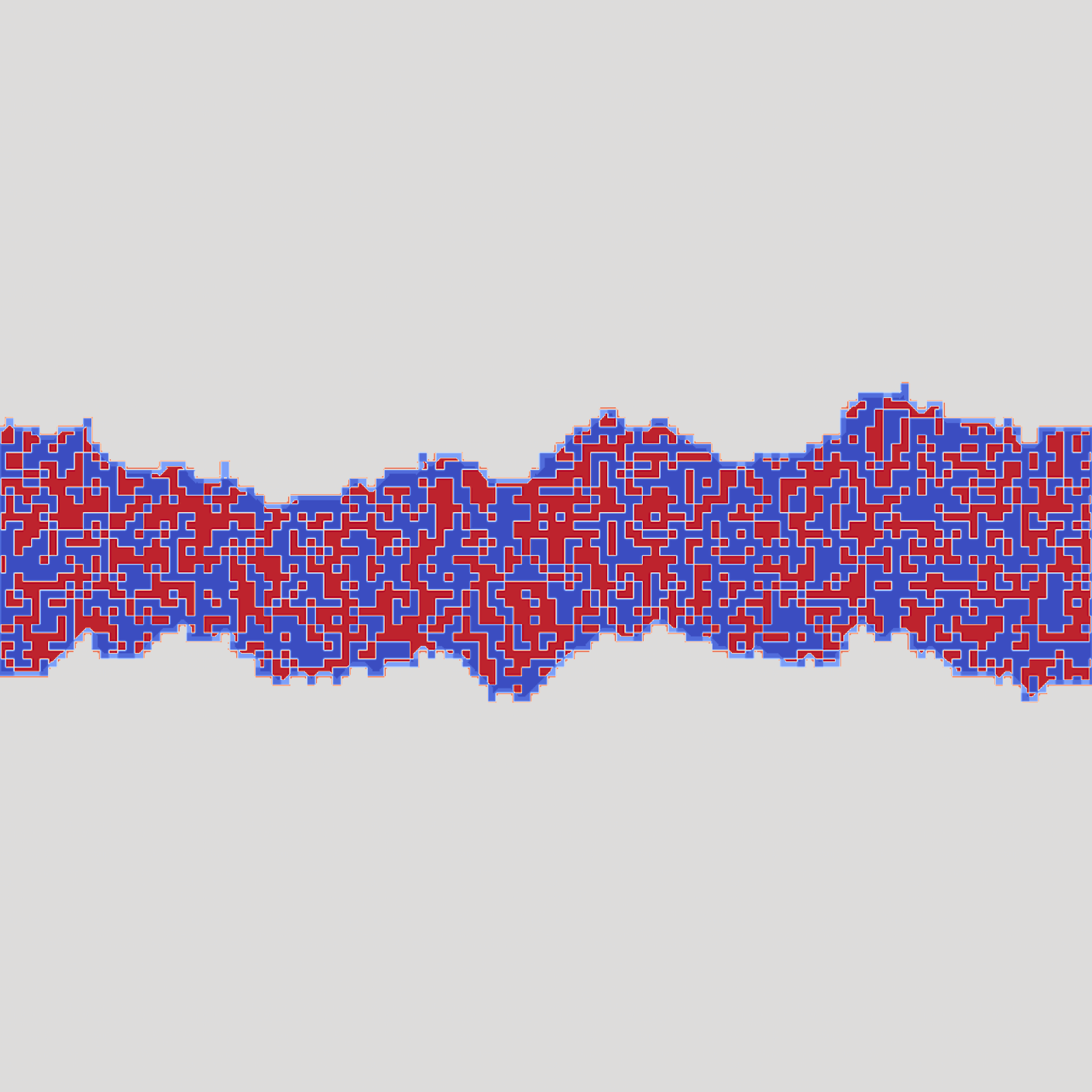}
        \centering
        \text{(d) Random}
    \end{minipage}
    \hfill
    \begin{minipage}{0.08\textwidth}
        \includegraphics[width=\textwidth]{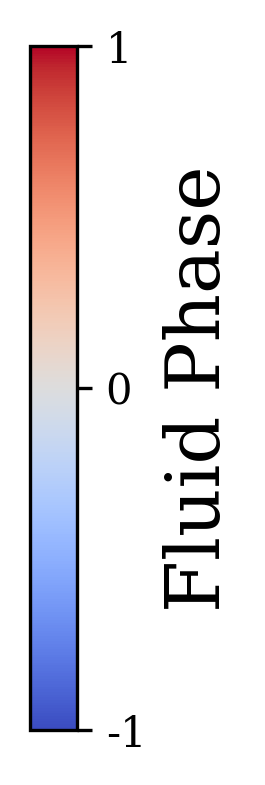}
        \centering
        \text{}
    \end{minipage}
    \caption{Comparison of fluid densities initialized by different methods within the simulation domain for multiphase simulations utilizing the Shan-Chen model. The Diffusion-Based Initialization synthesizes continuous values similar to the LBM solution. In contrast, the Euclidean Distance-Based and Random Initialization methods produce discrete phase representations. The color bar indicates a continuous range from -1 to 1, which is how the fluid phases are normalized for the machine learning workflow.}
    \label{fig:init_comp}
\end{figure}

We observed that as the training progressed, the model was able to provide solutions very close to the stable configuration of the LBM solver for the training data, as shown in Figure \ref{fig:subplot_comparison}. Upon completing the training, we explore the computational advantages of our hybrid approach in data not present in the training set. We generate initial configurations with the trained diffusion model for geometries unseen during training and then assess the number of iterations needed for convergence. To validate our strategy, we benchmark our diffusion-based initialization against two commonly used fluid initialization methods: Euclidean distance-based initialization and random initialization. Additionally, we include a benchmark using the simulation solution as the starting point for a new simulation. Figure \ref{fig:init_comp} presents examples of the different initializations within the simulation domain.

First, the \textit{Diffusion-Based Initialization} method uses a realization from our trained model within a given geometry. Using the same wetting/non-wetting fluid ratio, the \textit{Euclidean Distance-Based Initialization} identifies the non-wetting region in the flow path based on provided saturations and positions the non-wetting fluid away from the solid boundary. This is based on the understanding that the invading fluid tends to occupy pores distant from solid walls \cite{blunt2017multiphase}. The \textit{Random Initialization} method randomly places fluid particles throughout the connected domain until the desired saturation is reached. Finally, The \textit{Simulation Solution-Based Initialization} uses converged fluid configurations from the LBM simulation as its baseline. It is important to understand that this method sets a benchmark that other initializations are not expected to exceed. We purposely do not use the velocity field saved from the previous simulation since our diffusion model is only trained to provide the fluid configuration, not the velocity field. Therefore, initializing a simulation with only the previously converged fluid flow and not the velocity field may still require several iterations to reconverge. This ensures fairness when comparing it with other methods.

A notable advantage of the Diffusion-Based Initialization is the capability to initiate continuous density values. Although we are mainly interested in defining the two phases distinctly, slight variations in their density values inform the LBM simulation about local capillary pressure gradients, among other physical features. In contrast, the Euclidean Distance-Based and Random Initialization methods can only produce discrete phases (see Figure \ref{fig:init_comp}). Thus, the simulation solution-based initialization method and the random-based initialization method represent the lower and upper bounds of convergence time, respectively, while the Euclidean-based initialization is a reasonable approach, grounded in the domain knowledge that invading fluid tends to occupy pores distant from solid walls \cite{blunt2017multiphase}.

Table \ref{tab:initialization_comparison} provides a detailed comparison of 100 simulation results on geometries not seen during training. We compare each of the initialization methods described above.

The simulation solution initialization method, although it may seem somewhat impractical, as it essentially uses the solution of a previous simulation, represents the best possible scenario for initialization. By doing so, it sets a clear benchmark for the fastest possible convergence. This approach allows us to understand the upper limits of efficiency and provides a standard against which the performance of other initialization methods can be measured. On average the simulation took $4.38 \times 10^4$  iterations to convergence,  an average simulation time of 8.5 minutes. These values represent the potential lower bound of computational demand for the diffusion model under optimal conditions. 

Once this benchmark is established, the diffusion-based initialization method shows an improvement in efficiency compared to other common multiphase simulation initialization methods. On average, the simulations using diffusion-based initialization take $9.61 \times 10^4$ iterations, roughly 20.8 minutes. The diffusion-based initialization reduces the number of iterations required by 1.3 and 2.4 times compared to the Euclidean and Random initializations, respectively over the entire test set. Even after including the diffusion generation time, our proposed hybrid model remains much faster than the commonly used initialization methods. These results offer compelling evidence of the efficacy of the approach. We hypothesize that the computational cost improvements could be even more pronounced in more complex scenarios, such as in 3D porous media.

\begin{table}[h]
\centering
\caption{Comparison of average number of iterations and simulation time for different initialization methods.}
\begin{tabular}{lcc}
\hline
\textbf{Initialization Method} & \textbf{Number of Iterations} & \textbf{Simulation Time (min)} \\
\hline
Simulation Solution           & 43,800  & 8.5  \\
Generative Diffusion & 96,100 & 20.8  \\
Euclidean  Initialization       & 123,500 & 26.7  \\
Random   Initialization         & 225,900  & 46.0  \\
\hline
\end{tabular}
\label{tab:initialization_comparison}
\end{table}


\subsection{Testing on Real Fracture Dataset}

\begin{figure}[h]
\centering
\includegraphics[width=1.0\textwidth]{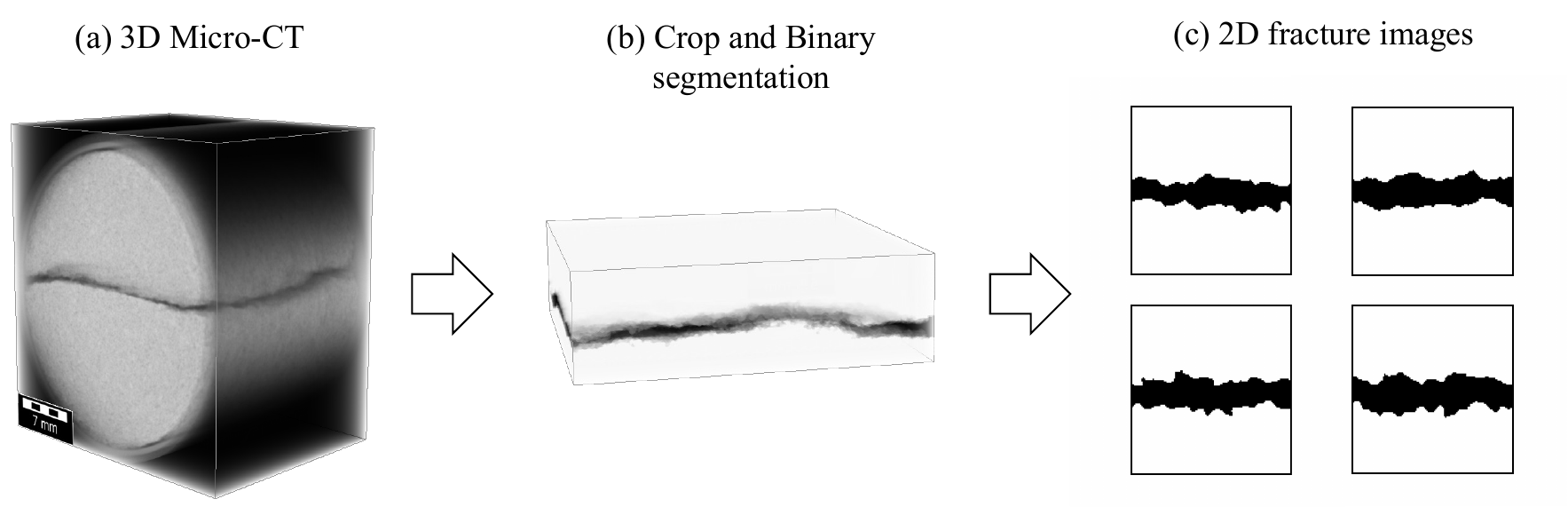}
\caption{Workflow for extracting 2D fracture images from the micro-CT scans of Berea sandstone \cite{dataset}. The process includes (a) reading in the large micro-CT file, (b) cropping and performing a binary segmentation on the geometry, and (c) extracting 2D fracture slices.}

\label{fig:3D-Micro_CT_2DFractures}
\end{figure}

\begin{figure}[h]
    \centering
    \begin{minipage}{0.22\textwidth}
        \includegraphics[width=\textwidth]{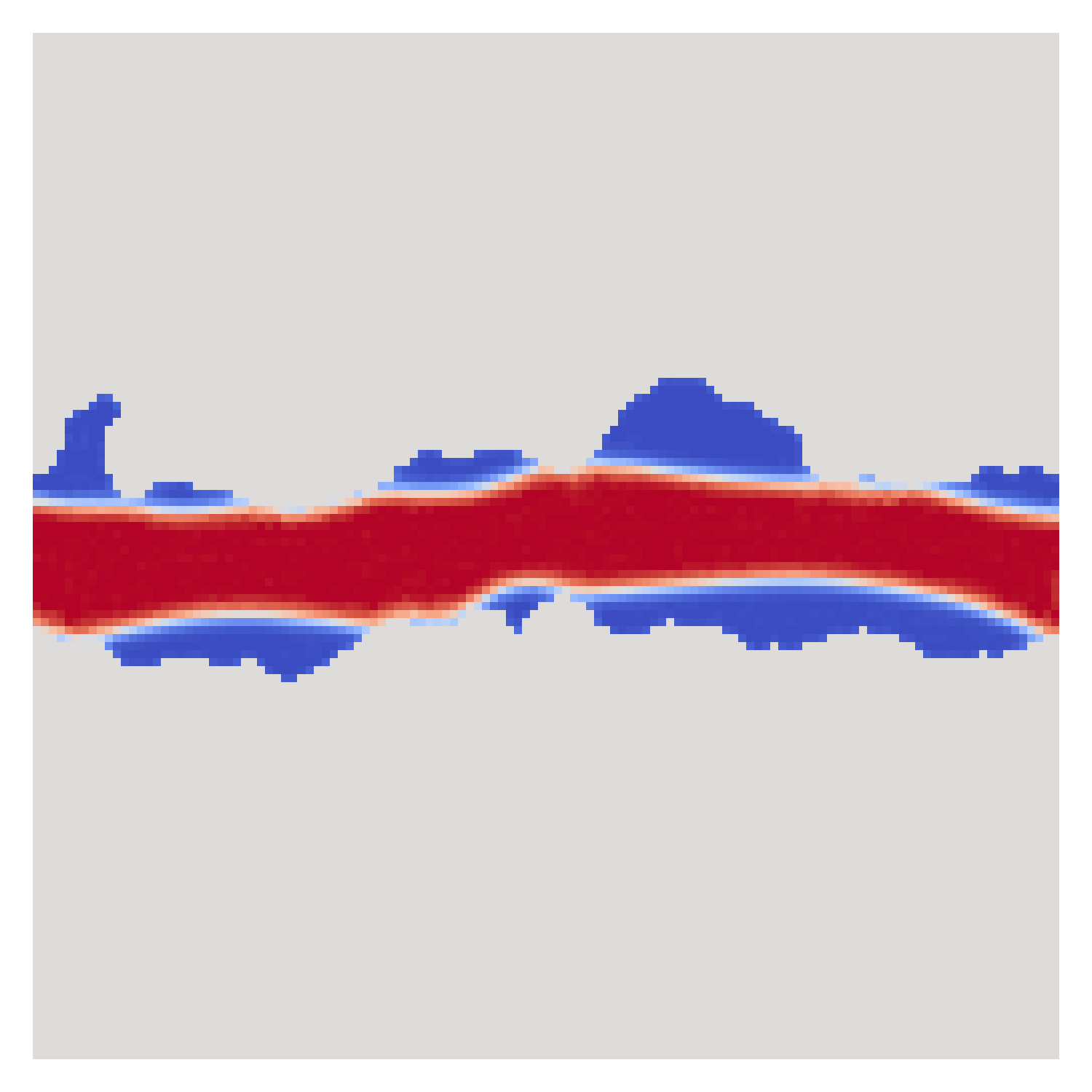}
        \centering
    \end{minipage}
    \hfill
    \begin{minipage}{0.22\textwidth}
        \includegraphics[width=\textwidth]{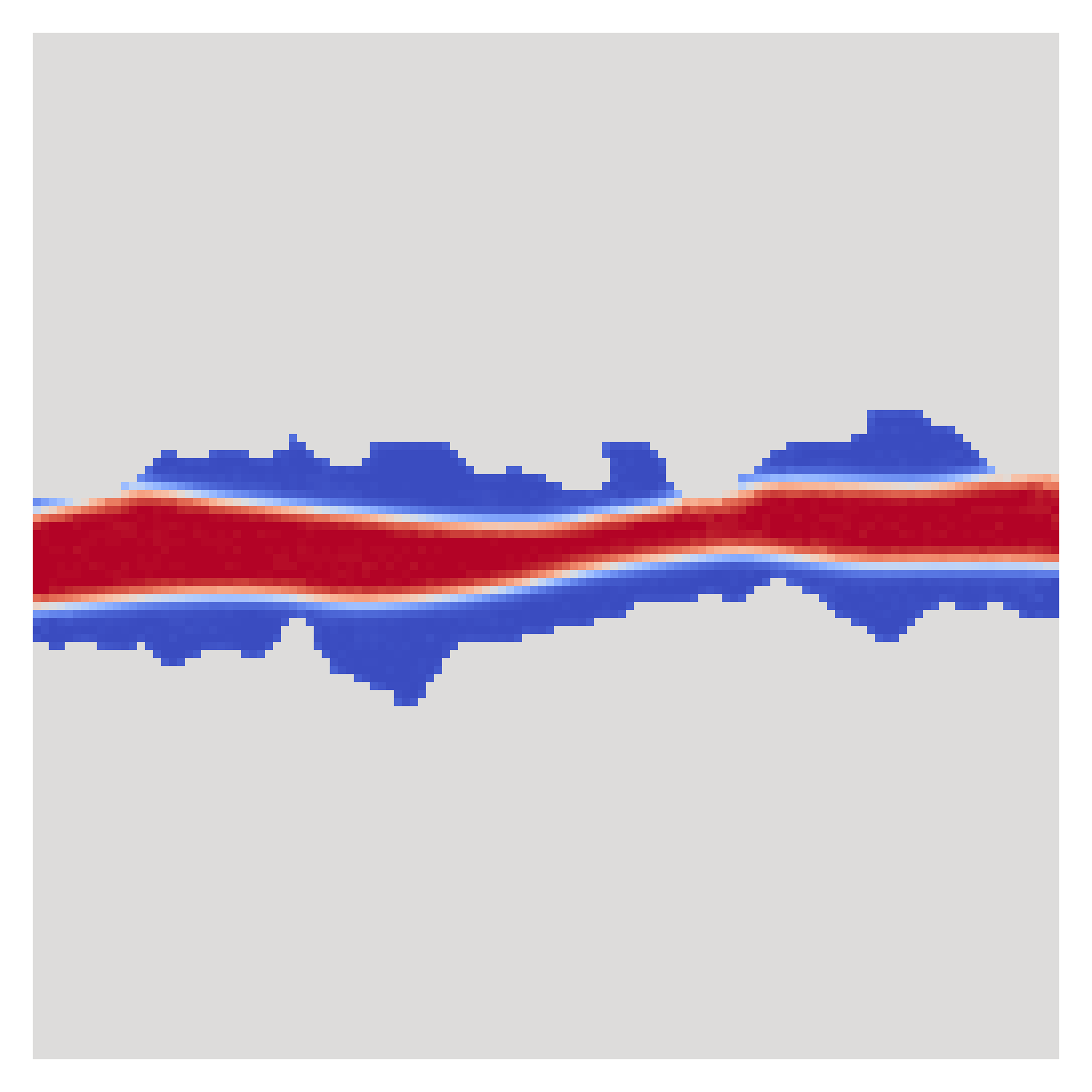}
        \centering
    \end{minipage}
    \hfill
    \begin{minipage}{0.22\textwidth}
        \includegraphics[width=\textwidth]{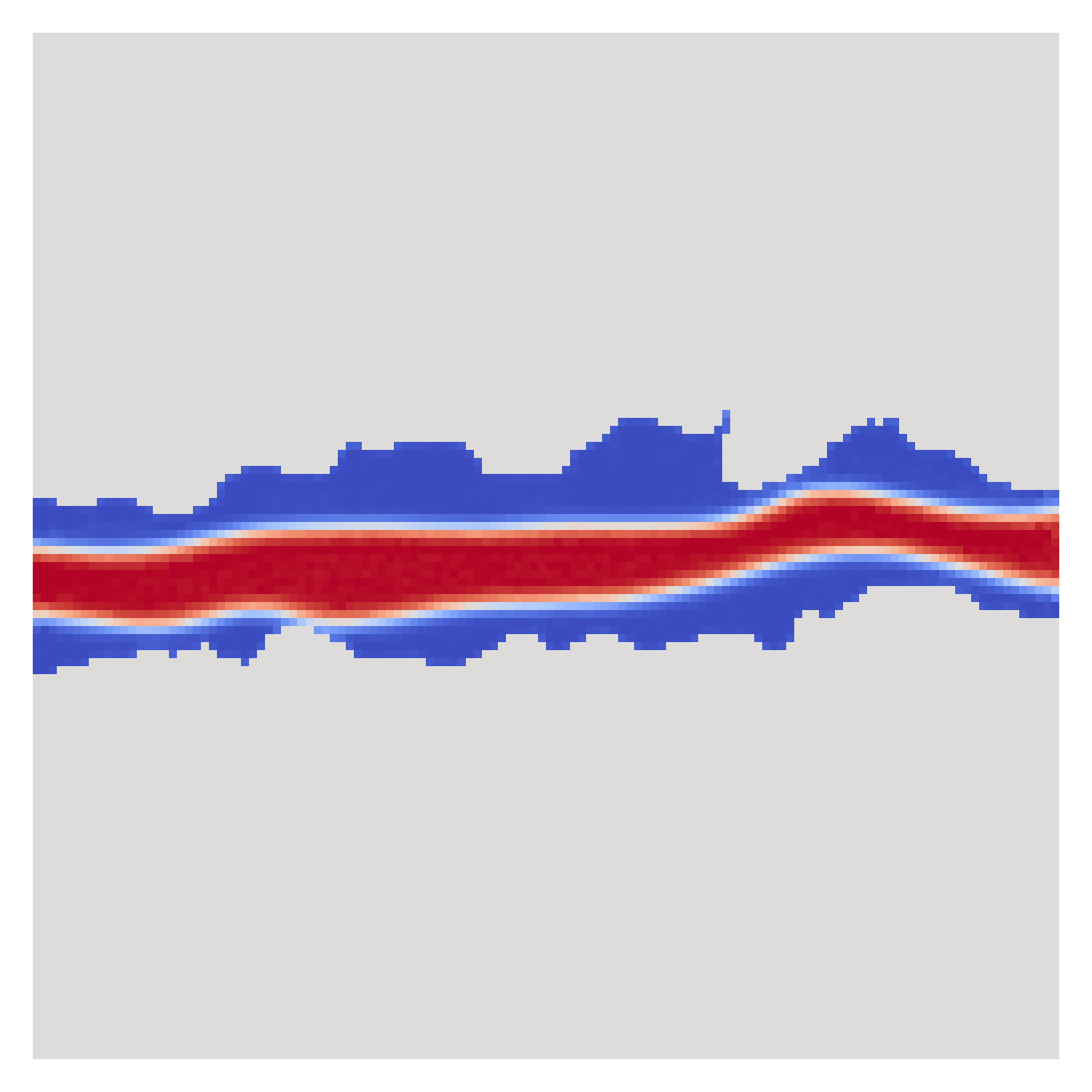}
        \centering
    \end{minipage}
    \hfill
    \begin{minipage}{0.22\textwidth}
        \includegraphics[width=\textwidth]{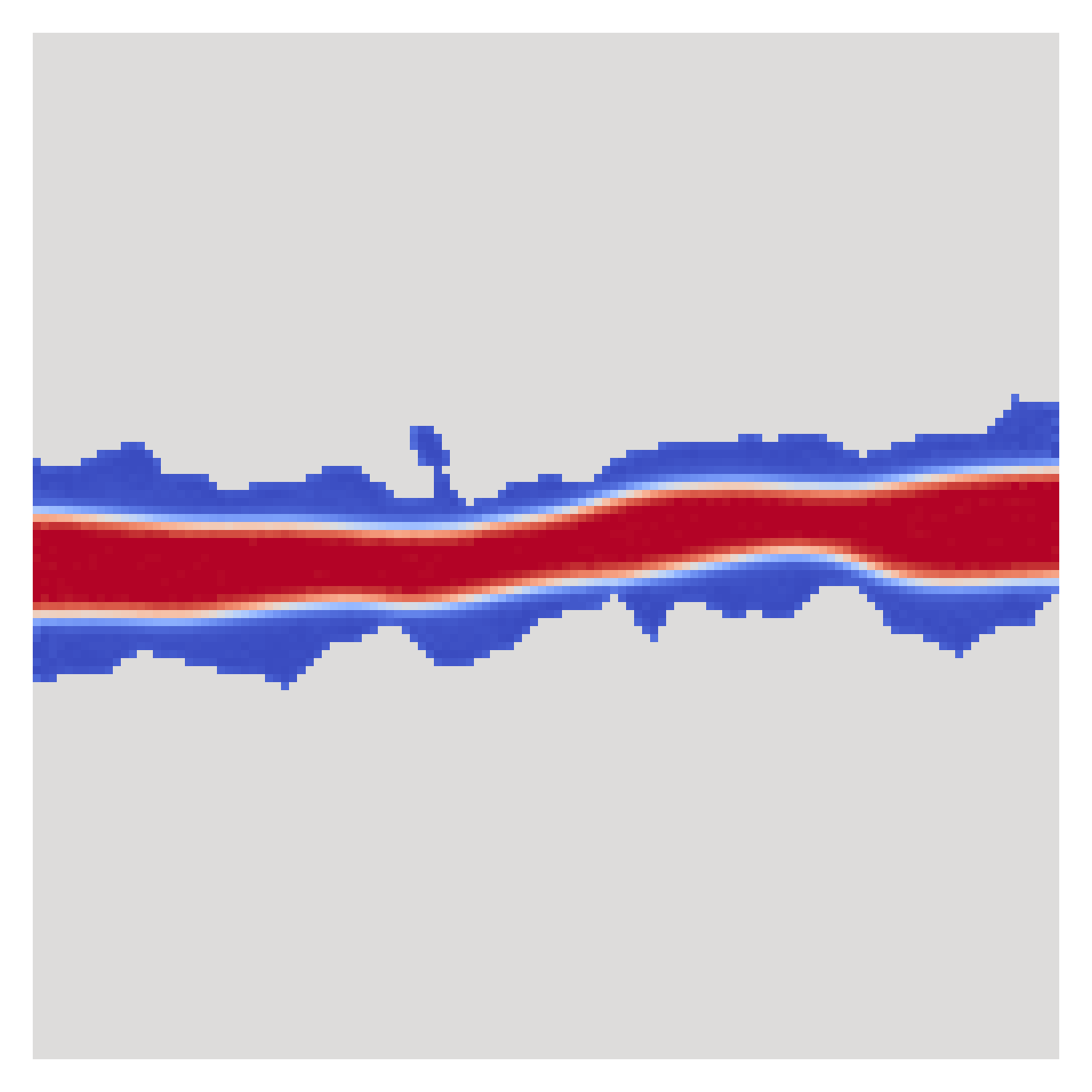}
        \centering
    \end{minipage}
    \hfill
    \begin{minipage}{0.08\textwidth}
        \includegraphics[width=\textwidth]{figs/colorbar_fluid_phase.png}
        \centering
    \end{minipage}    
    \caption{ Examples of generated multiphase fluid configurations in 4 different Berea fracture cross-sections.}
    \label{fig:generations_in_real_fractures}
\end{figure}

\begin{figure}[h]
\centering
\includegraphics[width=0.7\textwidth]{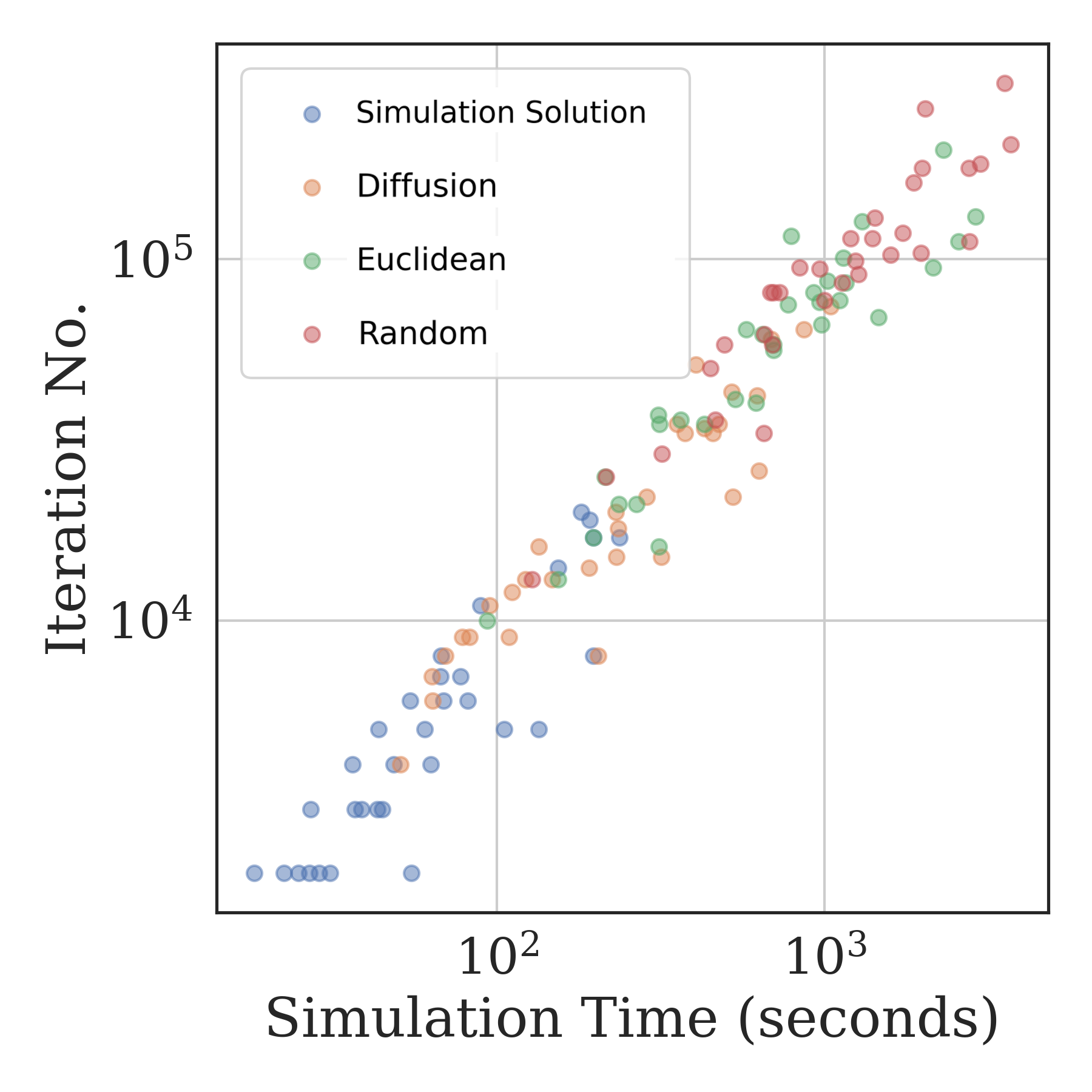}
\caption{Scatter plot comparing the number of iterations needed for convergence with the corresponding simulation times across different initialization methods for the real fracture dataset. Each point represents an individual LBM simulation, the average statistics for each initialization method is listed in Table\ref{tab:initialization_comparison_real_fracture}. }
\label{fig:Iter_vs_Time_real}
\end{figure}

We tested our trained model on a publicly-available real fracture in Berea sandstone \cite{karpyn2007visualization} with a voxel size of 27.344, 27.344, 32.548 $\mu$m.
We employed 2D fractures which are slices of the 3D fracture in the direction perpendicular to the longest coordinate of the image.
Despite training our model using simple synthetic fractures, we were able to accelerate the LBM simulations on the Berea fracture of 2.7 and 4.4 times compared to the Euclidean and Random initializations, respectively. Figure \ref{fig:3D-Micro_CT_2DFractures} shows the procedure to extract the fracture cross-sections from the 3D CT-scan \cite{dataset}. We then generated fluid realization on these cross-sections, some of these are shown in Figure \ref{fig:generations_in_real_fractures}. 
These configurations are then used as the starting points of the LBM simulations. In Figure \ref{fig:Iter_vs_Time_real} we show the number of iterations required for convergence and the simulation time for each initialization method, assessed on a dataset of 32 fractures, the average values for each initialization is reported in Table \ref{tab:initialization_comparison_real_fracture}. Consistent with the synthetic case, the computational costs in terms of iterations to convergence and simulation times still exhibit a clear advantage of our diffusion powered model over commonly used initializations for multiphase flow.

\begin{table}[h]
\centering
\caption{Berea fracture test results. Average number of iterations and average simulation time for 32 fracture geometries using different initialization methods.}
\begin{tabular}{lcc}
\hline
\textbf{Initialization Method} & \textbf{Number of Iterations } & \textbf{Simulation Time (min)} \\
\hline
Simulation Solution                & 6,500  & 1.3  \\
Generative Diffusion                       & 24,400  & 5.3  \\
Euclidean Initialization                       & 65,200  & 14.7  \\
Random   Initialization                         & 107,000  & 22.7 \\
\hline
\end{tabular}
\label{tab:initialization_comparison_real_fracture}
\end{table}

\section{Conclusion}
\label{sec:Conclusion}

In this study, we developed a hybrid approach that combines generative diffusion models with multiphase flow simulations. This method efficiently creates fluid configurations specific to fracture geometries, effectively managing varying saturation levels and significantly reducing the computational cost of pore-scale simulations. Our main contributions are the following. Fracture geometry conditioned training: Our model can generate fluid configurations within specific fracture geometries by incorporating geometric conditioning. This allows the model to accurately represent complex fracture surfaces and consider phenomena that occur at or near these interfaces, as well as in more distant regions. Simulation feedback during training: We introduced a new metric to evaluate our model's performance—the number of iterations needed for simulation convergence. This metric improved as the model trained, demonstrating its ability to understand both the visual and physical aspects of fluid configurations. It also provided a more meaningful stopping criterion for training compared to just using the loss function.

Thanks to these advancements and the ability of diffusion models to parameterize complex distributions, our model, although trained with relatively simple synthetic data, demonstrated its effectiveness on a real Berea sandstone fracture. The results showed that our model could generalize well to complex fractures, capturing the main factors affecting multiphase flow. Our hybrid method significantly reduced the iterations needed for simulation convergence, resulting in substantial computational savings. In the case of the Berea sandstone fracture, our model accelerated simulations by up a factor of 4.4 compared to commonly used initialization methods.

Overall, our study provides evidence that a diffusion model-based approach is effective for pore-scale simulations. Looking ahead, exploring three-dimensional (3D) simulations could be a valuable avenue for future research. However, creating a dataset for 3D simulations would be very computationally expensive, and adopting different neural networks would be necessary to manage the increased computational demands. Despite these challenges, the promising results from our study motivate further exploration in this direction. Models like this could provide researchers with a powerful tool to explore the sources of complex behaviors observed in nature. Furthermore, advancements in these algorithms could significantly transform reservoir modeling in the earth sciences. The future lies in hybrid approaches that combine machine learning with physics-based models, leveraging the strengths of both to achieve the best results. By integrating these methods, we can harness the computational efficiency of machine learning and the robustness of physics, demonstrating that both can coexist and complement each other effectively.

\section*{Open Research Section}
The code for generating synthetic fractures, multiphase flow simulations, and denoising diffusion models is available at the following repositories:
\begin{itemize}
    \item \textit{pySimFrac}: \href{https://github.com/lanl/pySimFrac}{https://github.com/lanl/pySimFrac}
    \item MP-LBM: \href{https://github.com/je-santos/MPLBM-UT}{https://github.com/je-santos/MPLBM-UT}
    \item Diffusion Model Codebase: \href{https://github.com/lucidrains/denoising-diffusion-pytorch}{https://github.com/lucidrains/denoising-diffusion-pytorch}
\end{itemize}


\section*{Acknowledgments}
J.C. thanks the Applied Machine Learning (AML) summer program and Center for Nonlinear Studies (CNLS) at Los Alamos National Laboratory (LANL) for the mentoring and the fellowship received. AM gratefully acknowledge the support of the Center for Non-Linear Studies (CNLS) for this work. J.E.S and Y.T. acknowledge the support by the Laboratory Directed Research \& Development project ``Diffusion Modeling with Physical Constraints for Scientific Data (20240074ER)''.

\bibliographystyle{plainnat.bst}
\bibliography{bibliography.bib}

\newpage
\appendix
\counterwithin{figure}{section}
\counterwithin{table}{section}

\end{document}